\documentclass[prd,superscriptaddress,nofootinbib,amsmath,amssymb,aps,11pt]{revtex4}

\usepackage{bm}
\usepackage{amsfonts}
\usepackage{latexsym}
\usepackage[latin1]{inputenc}
\usepackage{graphicx}
\usepackage{amsmath}
\usepackage{xcolor}
\usepackage{palatino}
\usepackage{mathpazo}
\usepackage{booktabs}
\usepackage{dcolumn}

\def\jnl@style{\it}
\def\aaref@jnl#1{{\jnl@style#1}}

\def\aaref@jnl#1{{\jnl@style#1}}

\def\aj{\aaref@jnl{AJ}}                   % Astronomical Journal
\def\apj{\aaref@jnl{ApJ}}                 % Astrophysical Journal
\def\apjl{\aaref@jnl{ApJ}}                % Astrophysical Journal, Letters
\def\apjs{\aaref@jnl{ApJS}}               % Astrophysical Journal, Supplement
\def\apss{\aaref@jnl{Ap\&SS}}             % Astrophysics and Space Science
\def\aap{\aaref@jnl{A\&A}}                % Astronomy and Astrophysics
\def\aapr{\aaref@jnl{A\&A~Rev.}}          % Astronomy and Astrophysics Reviews
\def\aaps{\aaref@jnl{A\&AS}}              % Astronomy and Astrophysics, Supplement
\def\mnras{\aaref@jnl{Mon.~Not.~Roy.~Astron.~Soc.}}             % Monthly Notices of the RAS
\def\prd{\aaref@jnl{Phys.~Rev.~D}}        % Physical Review D
\def\prc{\aaref@jnl{Phys.~Rev.~C}}  % Physical Review C
\def\prl{\aaref@jnl{Phys.~Rev.~Lett.}}    % Physical Review Letters
\def\qjras{\aaref@jnl{QJRAS}}             % Quarterly Journal of the RAS
\def\skytel{\aaref@jnl{S\&T}}             % Sky and Telescope
\def\ssr{\aaref@jnl{Space~Sci.~Rev.}}     % Space Science Reviews
\def\zap{\aaref@jnl{ZAp}}                 % Zeitschrift fuer Astrophysik
\def\nat{\aaref@jnl{Nature}}              % Nature
\def\aplett{\aaref@jnl{Astrophys.~Lett.}} % Astrophysics Letters
\def\apspr{\aaref@jnl{Astrophys.~Space~Phys.~Res.}} % Astrophysics Space Physics Research
\def\physrep{\aaref@jnl{Phys.~Rep.}}      % Physics Reports
\def\physscr{\aaref@jnl{Phys.~Scr}}       % Physica Scripta
\def\commat{\aaref@jnl{Comm.~Math.~Phys.}}              % Communications in Mathematical Physics
\def\science{\aaref@jnl{Science}}               % Science
\def\cqg{\aaref@jnl{Classical Quant.~Grav.}}            % Classical and Quantum Gravity
\def\jpcs{\aaref@jnl{JPCS}}                                     % Journal of Physics Conference Series
\def\ijmpd{\aaref@jnl{Int.~J.~Mod.~Phys.~D}}                    % International Journal of Modern Physics D
\def\grg{\aaref@jnl{Gen.~Relat.~Gravit.}}               % General Relativity and Gravitation
\def\rpp{\aaref@jnl{Rep.~Prog.~Phys.}}          % Reports on Progress in Physics
\def\npa{\aaref@jnl{Nucl.~Phys.~A}}        % Nuclear Physics A
\def\lrr{\aaref@jnl{Living Rev.~Rel.}}                   % Living reviews in relativity
\def\jcap{\aaref@jnl{J.~Cosmology Astropart.~Phys.}}    % Journal of cosmology and astroparticle physics
\def\rmp{\aaref@jnl{Rev.~Mod.~Phys.}}   %Reviews of modern physics

\definecolor{darkgreen}{rgb}{0.0, 0.8, 0.0}

\allowdisplaybreaks[1]
\begin{document}

	\title{Spin-induced scalarization of Kerr black holes with a massive scalar field}
	
	\author{Daniela D. Doneva}
	\email{daniela.doneva@uni-tuebingen.de}
	\affiliation{Theoretical Astrophysics, Eberhard Karls University of T\"ubingen, T\"ubingen 72076, Germany}
	\affiliation{INRNE - Bulgarian Academy of Sciences, 1784  Sofia, Bulgaria}
	
	\author{Lucas G. Collodel}
	\email{lucas.gardai-collodel@uni-tuebingen.de}
	\affiliation{Theoretical Astrophysics, Eberhard Karls University of T\"ubingen, T\"ubingen 72076, Germany}

	\author{Christian J. Kr\"uger}
	\email{christian.krueger@tat.uni-tuebingen.de}
	\affiliation{Theoretical Astrophysics, Eberhard Karls University of T\"ubingen, T\"ubingen 72076, Germany}

	\author{Stoytcho S. Yazadjiev}
	\email{yazad@phys.uni-sofia.bg}
	\affiliation{Theoretical Astrophysics, Eberhard Karls University of T\"ubingen, T\"ubingen 72076, Germany}
	\affiliation{Department of Theoretical Physics, Faculty of Physics, Sofia University, Sofia 1164, Bulgaria}
	\affiliation{Institute of Mathematics and Informatics, 	Bulgarian Academy of Sciences, 	Acad. G. Bonchev St. 8, Sofia 1113, Bulgaria}

	%%%%%%%%%%%%%%%%%%%%%%%%%%%%%%%%%%%%  DATE  %%%%%%%%%%%%%%%%%%%%%%%%%%%%%%%%%%%%
	%\date{\today}

	\begin{abstract}
		In the present paper we study the onset of the spin-induced scalarization of a Kerr black hole in scalar-Gauss-Bonnet gravity with a massive scalar field. Our approach is based on a (2+1) time evolution of  the relevant  linearized scalar field perturbation equation. We examine the region where the Kerr black hole becomes unstable giving rise to new scalarized rotating black holes with a massive scalar field. With increasing of the scalar field mass, the minimum value of the Gauss-Bonnet coupling parameter at which scalarization is possible, increases and thus the instability region shrinks. Interestingly, the introduction of scalar field mass does not change the critical minimal value of the black hole angular momentum $a_{\rm crit}/M$ where the instability of the Kerr black hole develops. 
	\end{abstract}
	
	\maketitle
	
	\section{Introduction}

	Spontaneous scalarization is a very interesting phenomenon that allows to endow the compact objects with scalar hair without altering the predictions in the weak field limit. The black hole scalarization is of particular interest and its study goes back to \cite{Stefanov2008,Doneva2010}. More recently, the 	black hole spontaneous scalarization was also discovered within the framework of Einstein-scalar-Gauss-Bonnet (EsGB) gravity \cite{PhysRevLett.120.131103,PhysRevLett.120.131104}. In simple words, the essence of spontaneous scalarization within Gauss-Bonnet gravity consists of the following: The standard general relativistic black holes, which are also solutions to the equations of the bigger theory, become unstable beyond a certain threshold in curvature and the black holes can acquire a nontrivial scalar hair. In order for the scalarization to occur, the coupling function $f(\varphi)$, describing the interaction between the scalar field and the spacetime curvature through the Gauss-Bonnet  scalar ${\cal R}^2_{GB}$, has to satisfy $\frac{df}{d\varphi}(0)=0$  and $\epsilon {\cal R}^2_{GB}<0$
	where $\epsilon=\frac{d^2 f}{d^2\varphi}(0)$. Beyond a certain threshold, the term  $\epsilon {\cal R}^2_{GB}<0$ in the scalar perturbation equation triggers the so-called tachyonic instability and the black holes develop scalar hair.  This was first demonstrated in the case $\epsilon>0$
	for Schwarzschild black holes \cite{PhysRevLett.120.131103, PhysRevLett.120.131104} and later for the Kerr solution \cite{PhysRevLett.123.011101,Collodel_2020}. Very recently, it was shown that the same occurs 
	in the case $\epsilon<0$---above a certain critical value of the angular momentum per unit mass, the Kerr black hole becomes unstable and
	this will eventually lead to formation of nontrivial scalar hair \cite{Dima:2020yac,Doneva:2020nbb,Hod:2020jjy}. The scalarization in the case $\epsilon<0$ was dubbed {\it spin-induced scalarization} since  the Schwarzschild black hole can not scalarize for $\epsilon<0$. 
	
	The purpose of the present paper is to study the spin-induced spontaneous scalarization in the case when the scalar field is massive.  The scalarization with a massive scalar field for $\epsilon>0$ was studied \cite{PhysRevD.99.104041,PhysRevD.99.104045,Macedo:2020tbm}.
	The inclusion of a mass term for the scalar field, or more generally self-interaction for the scalar field, is consistent with the principles of
	effective field theory  \cite{PhysRevD.99.104041}. However, only the mass term can alter the onset of the spontaneous scalarization while the self interaction affects the nonlinear effects in the scalarized solution. That is why  
	in the present paper we consider only the mass term. From a physical point of view, the inclusion of scalar field mass can change the picture considerably. It suppresses the scalar field at a length scale of the order of the Compton wavelength  of the scalar field which helps us reconcile the theory 	with the observations for a much broader range of the coupling parameters and functions \cite{PhysRevD.99.104045}. In addition, the scalar field mass shifts the bifurcation points of the scalarization, i.e. the points where new scalarized solutions branch out of the general relativistic one; in other words, it changes the threshold beyond which scalar hair develops similar to the case of the scalarization of neutron stars with a massive scalar field \cite{Staykov:2018hhc}.       
	
	In this paper we study the instability of Kerr EsGB black holes by fully  evolving in $2+1$ dimensions the modified Klein-Gordon  equation 
	describing the perturbation of the massive scalar field of Kerr black holes within EsGB gravity. The final goal is to determine in which regions of the parameter space that defines the theory a tachyonic instability gives rise to new hairy black holes.

	\section{Scalar field perturbations within Gauss-Bonnet gravity}	
	
	The  EsGB gravity is defined by the action   
	
	\begin{align}
	S & = \frac{1}{16\pi}\int d^4x \sqrt{-g} 
	\Big[R - 2\nabla_\mu \varphi \nabla^\mu \varphi - V(\varphi) 
	+ \lambda^2 f(\varphi){\cal R}^2_{GB} \Big] ,\label{eq:quadratic}
	\end{align}
	where $R$ is the Ricci scalar with respect to the spacetime metric $g_{\mu\nu}$, $\varphi$ is the scalar field with a potential $V(\varphi)$ and a coupling function  $f(\varphi)$ depending only on $\varphi$, $\lambda$ is the Gauss-Bonnet coupling constant having  dimension of $length$ and ${\cal R}^2_{GB}$ is the Gauss-Bonnet invariant\footnote{The Gauss-Bonnet invariant is defined by ${\cal R}^2_{GB}=R^2 - 4 R_{\mu\nu} R^{\mu\nu} + R_{\mu\nu\alpha\beta}R^{\mu\nu\alpha\beta}$, where $R$ is the Ricci scalar, $R_{\mu\nu}$ is the Ricci tensor and $R_{\mu\nu\alpha\beta}$ is the Riemann tensor}. The above action yields the following field equations 
	
	\begin{align}\label{FE}
	R_{\mu\nu}- \frac{1}{2}R g_{\mu\nu} + \Gamma_{\mu\nu}
	& = 2\nabla_\mu\varphi\nabla_\nu\varphi -  g_{\mu\nu} \nabla_\alpha\varphi \nabla^\alpha\varphi - \frac{1}{2} g_{\mu\nu}V(\varphi),\\
	\nabla_\alpha\nabla^\alpha\varphi
	& = \frac{1}{4} \frac{dV(\varphi)}{d\varphi} -  \frac{\lambda^2}{4} \frac{df(\varphi)}{d\varphi} {\cal R}^2_{GB},
	\end{align}
	where  $\nabla_{\mu}$ is the covariant derivative with respect to  $g_{\mu\nu}$, while  $\Gamma_{\mu\nu}$ is given by 
	
	\begin{align}
	\Gamma_{\mu\nu}
	& = - R(\nabla_\mu\Psi_{\nu} + \nabla_\nu\Psi_{\mu} ) - 4\nabla^\alpha\Psi_{\alpha}\left(R_{\mu\nu} - \frac{1}{2}R g_{\mu\nu}\right) + 
	4R_{\mu\alpha}\nabla^\alpha\Psi_{\nu} + 4R_{\nu\alpha}\nabla^\alpha\Psi_{\mu} \nonumber \\ 
	& \qquad - 4 g_{\mu\nu} R^{\alpha\beta}\nabla_\alpha\Psi_{\beta} 
	+ \,  4 R^{\beta}_{\;\mu\alpha\nu}\nabla^\alpha\Psi_{\beta} 
	\end{align}  
	
	with 
	
	\begin{align}
	\Psi_{\mu} & := \lambda^2 \frac{df(\varphi)}{d\varphi}\nabla_\mu\varphi .
	\end{align}
	
	In the present paper, we shall consider asymptotically flat spacetimes. Without loss of generality, 
	we can choose the asymptotic  value of the scalar field to be zero  and we can impose the following constraints on the 
	coupling function $f(\varphi)$: $f(0)=0$ and $\frac{d^2f}{d\varphi^2}(0)=\epsilon$ with $\epsilon=\pm 1$. 
	Since the focus of the present paper
	paper is on spontaneous scalarization, we impose one more condition on $f(\varphi)$, namely  $\frac{df}{d\varphi}(0)=0$, which is crucial for the spontaneous scalarization. Additionally, the asymptotic flatness imposes the following conditions on the potential $V(\varphi)$, namely
	$V(0)=\frac{dV}{d\varphi}(0)=0$. Under these conditions it is not difficult to see that the Kerr black hole solution is also 
	a solution to the EsGB gravity with a trivial scalar field, i.e.,  $\varphi=0$. 
	However, beyond a certain threshold in curvature, both static and rotating solutions could become unstable and acquire scalar hair for $\epsilon>0$  \cite{PhysRevLett.120.131103, PhysRevLett.120.131104}, while in the $\epsilon<0$ case, rapidly rotating black holes could suffer from a spin-induced tachyonic instability \cite{Dima:2020yac,Doneva:2020nbb}.  In order to determine where in the parameter space the onset of the scalarization occurs, we have to study the stability of the Kerr solution within the framework of the EsGB gravity with 
	a massive scalar field. 
	
	In order to study the stability of the Kerr black hole  we shall consider the perturbation of the Kerr solution within the framework of EsGB gravity. It is not difficult to see that when the condition $\frac{df}{d\varphi}(0)=V(0)=\frac{dV}{d\varphi}(0)=0$ is met, the equations governing the perturbations of the metric $\delta g_{\mu\nu}$ are decoupled from the equation governing the perturbation $\delta \varphi$ of the scalar field. The equations for metric perturbations  are in fact the same as those in the pure Einstein gravity and therefore we shall focus only on the scalar field perturbations. The equation for the scalar perturbation is
	
	\begin{eqnarray}\label{PESF}
	\Box_{(0)} \delta\varphi = \left(m_{\varphi}^2 -\frac{\epsilon}{4}\lambda^2  {\cal R}^2_{GB(0)}\right) \delta\varphi, 
	\end{eqnarray} 
	where $m^2_{\varphi}= \frac{1}{4}\frac{d^2 V}{d\varphi^2}(0)$ is square of the mass of the scalar field and  
	$\Box_{(0)}$ and ${\cal R}^2_{GB(0)}$ are the d'Alembert operator and the Gauss-Bonnet invariant for 
	Kerr geometry. Since the focus is on spin-induced scalarization we choose $\epsilon=-1$. Note that the $\delta\varphi$ coefficient on the right hand side of eq. \ref{PESF} above gives a non-homogeneous effective mass for the scalar field
	\begin{equation}
	\mu^2_{\rm eff} = m_{\varphi}^2 -\frac{\epsilon}{4}\lambda^2  {\cal R}^2_{GB(0)}
	\end{equation}
	and tachyonic instability can only occur if this term is not positive everywhere.
	
	The Kerr metric presented in the standard  Boyer-Lindquist coordinates reads   
	\begin{eqnarray}\label{KerrM}
	ds^2= - \frac{\Delta -a^2\sin^2\theta}{\Sigma} dt^2 - 2a \sin^2\theta \frac{r^2 + a^2 - \Delta}{\Sigma} dt d\phi  
	+ \frac{(r^2 + a^2)^2 - \Delta a^2 \sin^2\theta}{\Sigma} \sin^2\theta d\phi^2 + \frac{\Sigma}{\Delta} dr^2 + \Sigma d\theta^2
	\end{eqnarray} 
	where
	\begin{equation}
	\Delta := r^2 - 2Mr + a^2 \quad\text{and}\quad \Sigma := r^2 + a^2 \cos^2\theta.
	\end{equation}
	The Gauss-Bonnet invariant for the Kerr geometry can be written as 
	\begin{eqnarray}
	{\cal R}^2_{GB(0)}= \frac{48 M^2}{\Sigma^6}(r^2- a^2\cos^2\theta)(r^4 - 14 a^2 r^2 \cos^2\theta + a^4\cos^4\theta).
	\end{eqnarray} 
	
	In writing  the perturbation equation (\ref{PESF}) in explicit form  it is important to introduce a new azimuthal coordinate $\phi_*$ defined by 
	\begin{eqnarray}
	d\phi_*=d\phi + \frac{a}{\Delta} dr.  
	\end{eqnarray} 
	This new azimuthal coordinate helps us to get rid of some unphysical pathologies near the horizon. It is also convenient to use  the tortoise coordinate $x$ defined by 
	
	\begin{eqnarray}
	dx := \frac{r^2+a^2}{\Delta} dr.  
	\end{eqnarray} 
	
	In the new coordinates $(t,x,\theta,\phi_*)$, the scalar perturbation equation (\ref{PESF}) takes the following explicit form
	
	\begin{align} \label{eq:PertEq}
	& -\left[(r^2 + a^2)^2 - \Delta a^2 \sin^2\theta\right] \partial^2_t \delta\varphi + (r^2 + a^2)^2 \partial^2_x \delta\varphi + 2r \Delta \partial_x\delta\varphi - 4Ma r\partial_t\partial_{\phi_*}\delta\varphi \nonumber \\ 
	& +  2a(r^2 + a^2)\partial_x\partial_{\phi_*}\delta\varphi  + \Delta\left[\frac{1}{\sin\theta} \partial_\theta(\sin\theta\partial_\theta\delta\varphi) +   \frac{1}{\sin^2\theta}\partial^2_{\phi_*}\delta\varphi \right] \\ 
	& \quad = \left[m_\varphi^2 \Delta\Sigma      + \lambda^2 \frac{12 M^2\Delta}{\Sigma^5}(r^2- a^2\cos^2\theta)(r^4 - 14 a^2 r^2 \cos^2\theta + a^4\cos^4\theta)\right]\delta\varphi. \nonumber
	\end{align} 
	
	The boundary conditions we have to impose when evolving in time eq. \eqref{eq:PertEq} is that the scalar field perturbation has the form of an outgoing wave at infinity and an ingoing wave at the black hole horizon.
	
	\section{Numerical method}
	The perturbation equation  \eqref{eq:PertEq} is formally  a Klein-Gordon equation  with variable  effective  mass, as defined above, on the Kerr background.  Since the background is axisymmetric  we can assume the following form of the scalar field perturbation   
	\begin{equation}\label{eq:phi_ansatz}
	\delta \varphi (t,x,\theta,\phi_*) = \delta \varphi (t,x,\theta) e^{im\phi_*},
	\end{equation}
	where $m$ is an integer---the well-known azimuthal mode number. After substituting this definition into Eq. \eqref{eq:PertEq}, we will arrive at a perturbation equation in (2+1) dimensions, meaning two spatial dimensions in addition to the time dimension. In addition, $m$ will enter explicitly in the resulting equation.
	
	The (2+1) time evolution of similar types of perturbation equations was performed in \cite{Krivan:1996da,Krivan:1997hc,Zenginoglu:2010cq,Harms:2013ib,Harms:2014dqa} in the general relativistic case, and in \cite{Gao:2018acg} for instabilities in Chern-Simons gravity. We will follow the approach described in detail in \cite{Doneva:2020nbb}, where the spin-induced scalarization in the absence of scalar field mass was examined. As a matter of fact the approach is very similar to the evolution of spacetime perturbations around rotating neutron stars considered in  \cite{Kruger:2019zuz,Kruger:2020ykw}. 
	
	Details about the code implementation can be found in \cite{Doneva:2020nbb}. Here we will comment on the most important points. 
	The relevant perturbation equation \eqref{eq:PertEq}, after the substitution \eqref{eq:phi_ansatz}, is transformed to a system of four real equations that are first order in time and their integration is performed with a 3rd order Runge-Kutta method. The ingoing and outgoing boundary conditions at the black hole horizon and at infinity, respectively, are independent of the angular coordinate and they are imposed following \cite{RuoffPhD}. In this approach, though, there will inevitably be spurious reflection from the outer boundaries of the grid that can be ``cured'' (in the sense that it does not impact the observed signal for a sufficiently long evolution time) simply by pushing the right boundary to very large values. In this way the small reflected signal from infinity  needs a very long time to travel to the point of observation and practically does not influence the observed signal for the first few milliseconds. At the rotation axis  ($\theta = 0$ and $\theta = \pi$) we impose $\delta \varphi=0$ for $m>0$  and $\partial\delta\varphi/\partial\theta=0$ for $m = 0$.

	The initial condition we impose has the form of a Gaussian pulse in $x$ direction with zero velocity, located at $x=12$, having unit amplitude and width $\sigma=1$. In $\theta$ direction the pulse has the form of the spherical harmonic of order $l$. The perturbation equation \eqref{eq:PertEq} is practically independent of $l$, but it turns out that in most case (if the scalar field evolution is stable), predominantly the mode with the same $l$ as the initial data is excited. In the unstable case, the exponential growth will posses the features of the fastest growing mode for a fixed $m$, independent of the initial perturbation.
	
	\section{Results}
	
	\subsection{Scalar field perturbations for $m_\varphi \ne 0$ and spin-induced scalarization}
	
\begin{figure}[htp]
		\centering
		\includegraphics[width=0.45\textwidth]{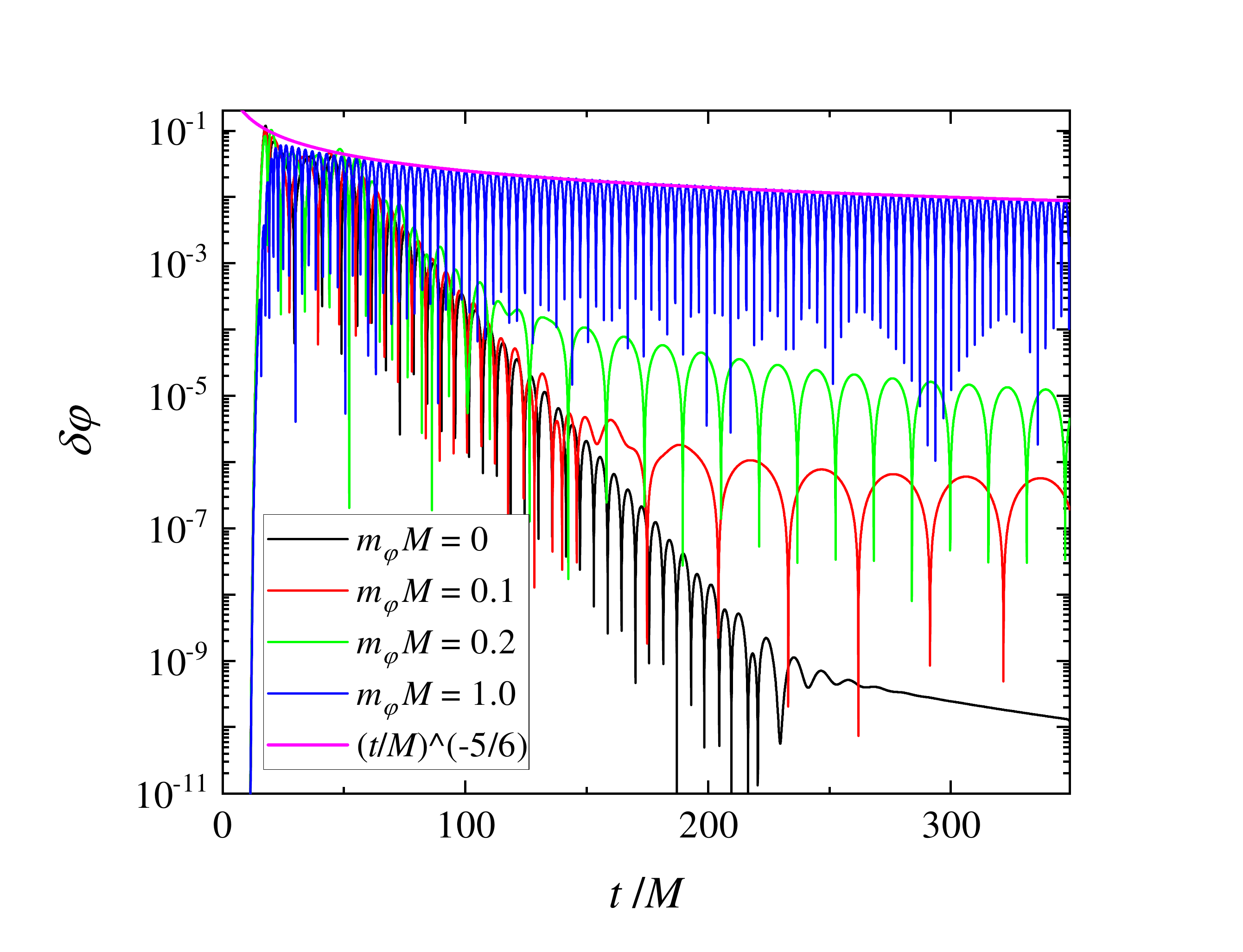}
		\includegraphics[width=0.45\textwidth]{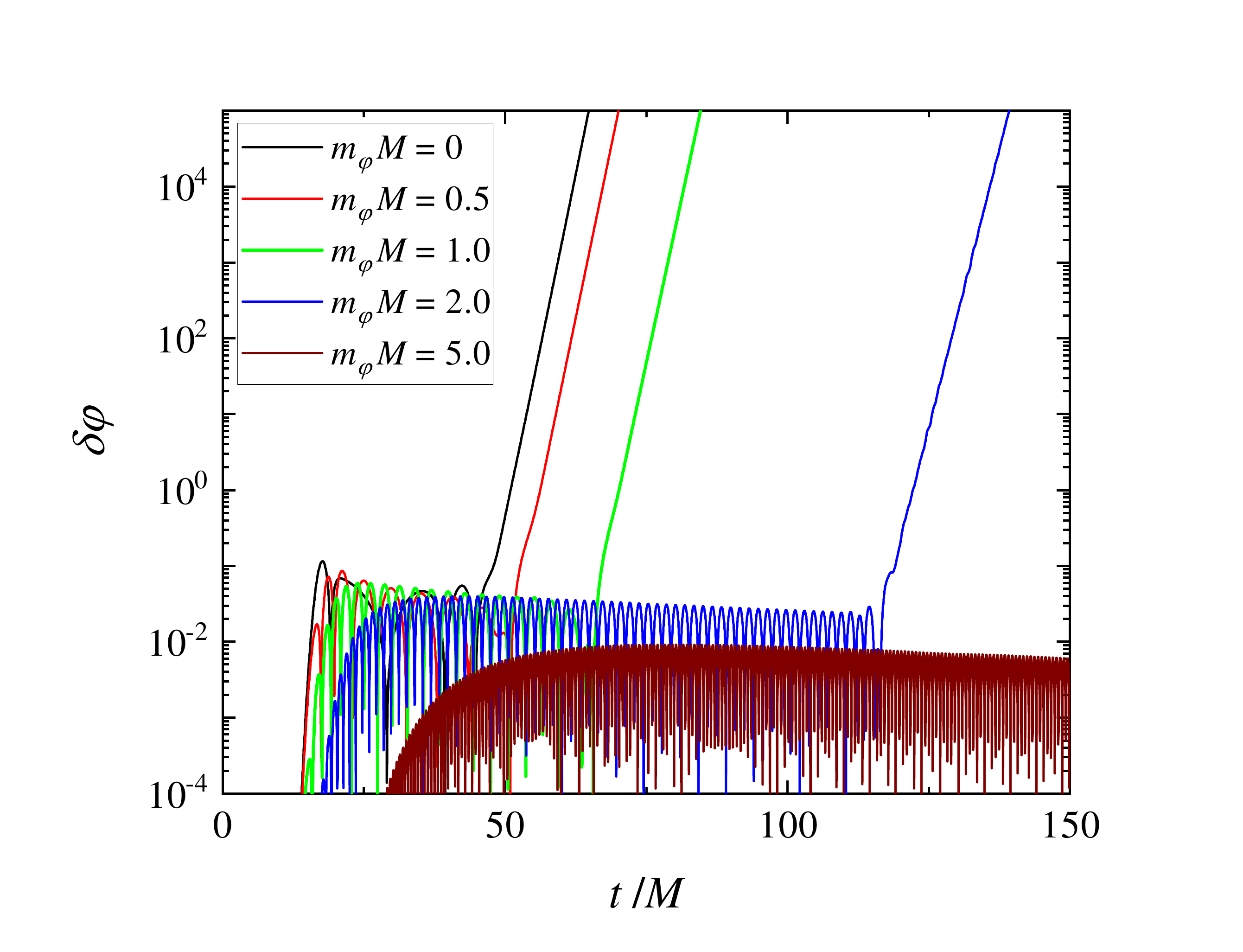}
		\caption{The time evolution of $\delta \varphi$ for some representative stable (left) and unstable (right) black holes. In the stable case we have chosen $a/M=0.4$, $\lambda/M=10$ and several values of $m_\varphi$, while for the unstable case we have chosen $a/M=0.8$ and $\lambda/M=100$. In the left panel, the expected analytical behavior of the late-time asymptotic of $\delta \varphi$ is plotted, i.e., $\delta \varphi \sim (t/M)^{-5/6}$.}
		\label{fig:time_evol}
	\end{figure}

	Let us first focus on the case of stable scalar quasi-normal modes (QNMs). In this case the time evolution of massive scalar field has certain specifics compared to the massless case. Fig. \ref{fig:time_evol} shows the signal observed at $x/M=30$ for both stable (left panel with $a/M=0.4$) and unstable (right panel with $a/M=0.8$) background models; the presented results are for $\lambda/M=10$ and several scalar field masses. The normalized scalar field mass $m_\varphi M$ is roughly the ratio between the horizon radius and  the Compton wavelength of the scalar field. The cases presented in the figure are representative examples and  the results are qualitatively similar when choosing other values for $a$, $\lambda$ and $m_\varphi$. For these simulations we have used $2000\times 60$ grid points in $x$ and $\theta$ direction. The computational domain spans from $x_{-\infty}=-20$ up to $x_{\infty}=200$, while $\theta \in [0,\pi]$.
	
	As one can see in the left panel of Fig. \ref{fig:time_evol}, the asymptotic tail appearing for $m_\varphi=0$ at late times is substituted by an oscillatory behavior when $m_\varphi>0$. The analytical behavior of this asymptotic tail at late times has the form \cite{Price:1971fb,Burko:2004jn} 
	\begin{equation}
	\delta \varphi_{\rm tail} \sim \cos(m_\varphi t) t^{-5/6}
	\end{equation} 
	and it can be shown that it is practically independent of the rotation rate and the  value of $\lambda$ (as long as we have a stable scalar field evolution). We have checked that indeed the frequency of the tail is in very good agreement with the analytically predicted value of $f_{\rm tail}=m_\varphi/2\pi$. In addition, the analytical function $t^{-5/6}$  is plotted in Fig. \ref{fig:time_evol} (suitably normalized); it is evident that it fits perfectly to the temporal decay of the time signal from the simulation with $m_\varphi M=1.0$. We have opted to fit the analytic function to the $m_\varphi M = 1.0$ curve purely for reasons of clarity; we can confirm that the other curves have the same power-law decay at late times.
	
	If the mass of the scalar field is small enough so that $f_{\rm tail}$ is lower than the QNM frequency of the scalar field perturbation, the signal is first dominated by the QNM ringing followed by the oscillatory tail. For larger $m_\varphi$, though, where $f_{\rm tail}$ is comparable or larger than the QNM frequency, the mass term dominates during the whole evolution. A general observation is that the oscillatory tail appears earlier in time with the increase of $m_\varphi$. 
	
	The behavior in the case of an unstable scalar field perturbation (right panel of Fig. \ref{fig:time_evol}) is somehow similar to the stable case before the onset of the instability. Once the exponential growth of $\delta \varphi$ starts, it turns out that the growth time $\tau$ of the mode, defined as $\delta \varphi \sim \exp(t/\tau)$ is only weakly dependent on the scalar field mass as one can see  in Fig. \ref{fig:time_evol} (right panel). From analytical considerations it is clear that the mass term suppresses the instability; this is indeed evident in the right panel of Fig. \ref{fig:time_evol} -- the onset of the exponential growth is shifted to later times with increasing mass $m_\varphi$ and eventually it ceases to exist for large enough $m_\varphi$ (the dark red line in the figure). Thus, for every fixed $a$ and $\lambda$ there is a threshold $m_\varphi$ that stabilizes the Kerr black hole.

	\subsection{Instability region of the Kerr black hole in massive EsGB gravity}
	
\begin{figure}[htp]
		\centering
		\includegraphics[width=1.1\textwidth]{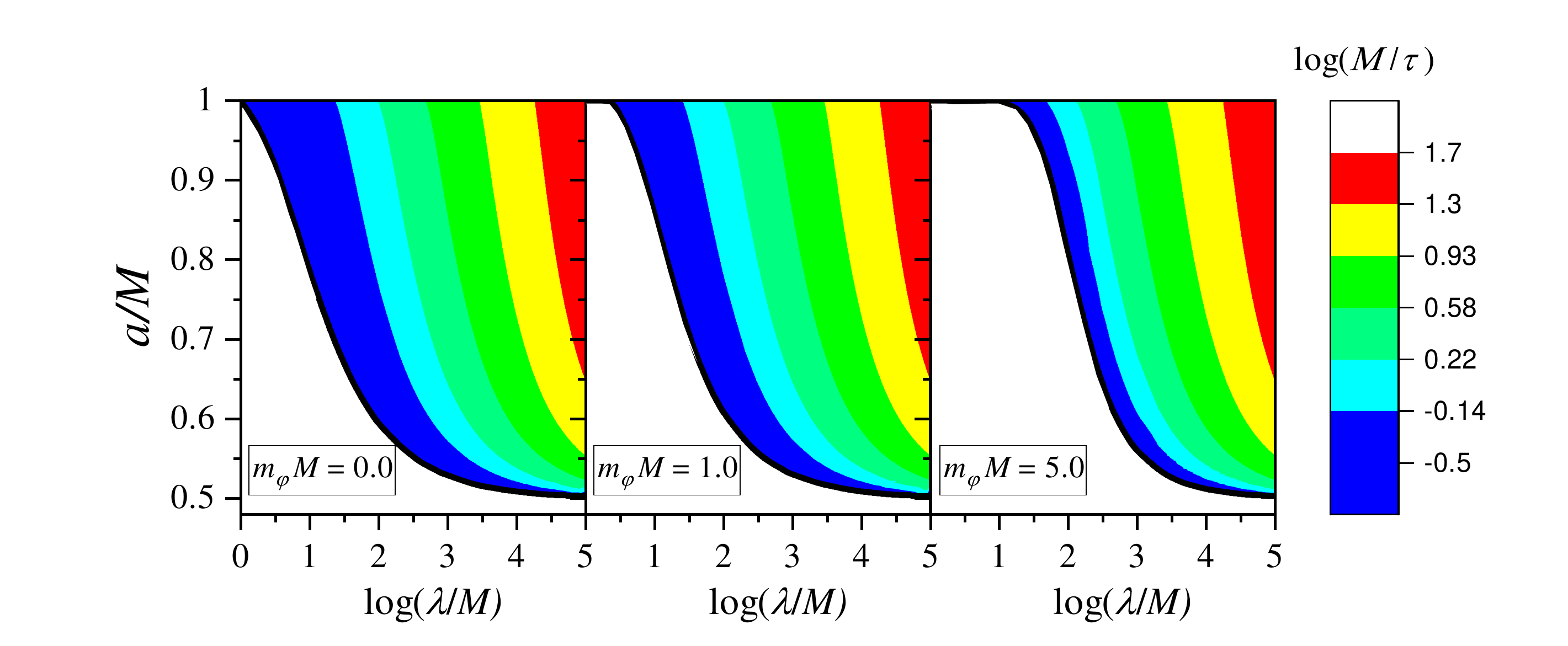}
		\caption{Contour plots of the growth time of the unstable modes for $m=0$ and several values of the scalar field mass -- $m_\varphi M=0$ (\textit{left}), $m_\varphi M=1.0$ (\textit{middle}), and $m_\varphi M=5.0$ (\textit{right}). In addition, the threshold lines, i.e. the contour lines $a(\lambda)$ dividing the parameter space into stable and unstable regions, are plotted with solid black lines in each plot for the corresponding $m_\varphi$.}
		\label{fig:contour}
	\end{figure}

\begin{figure}[htp]
		\centering
		\includegraphics[width=0.45\textwidth]{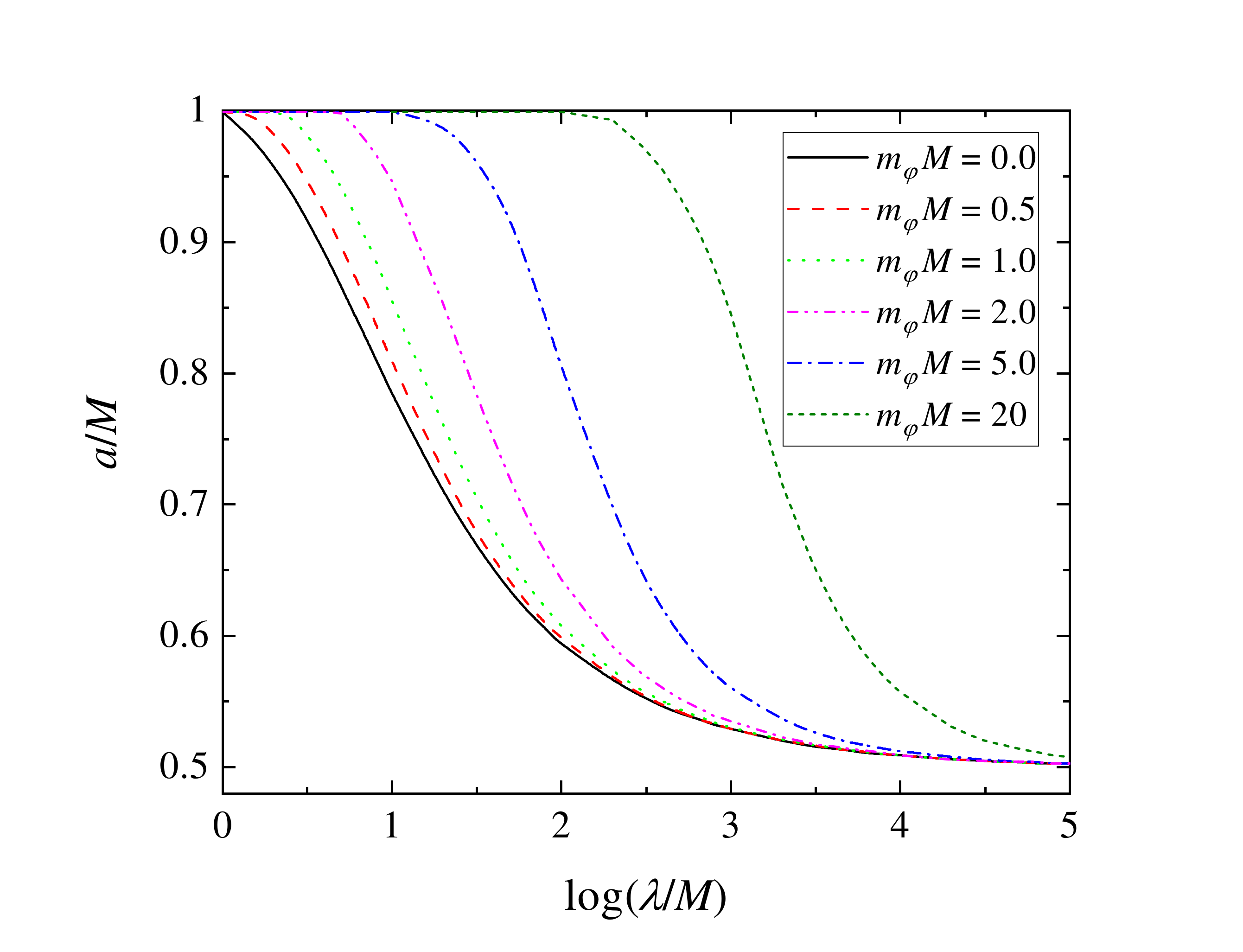}
		\includegraphics[width=0.45\textwidth]{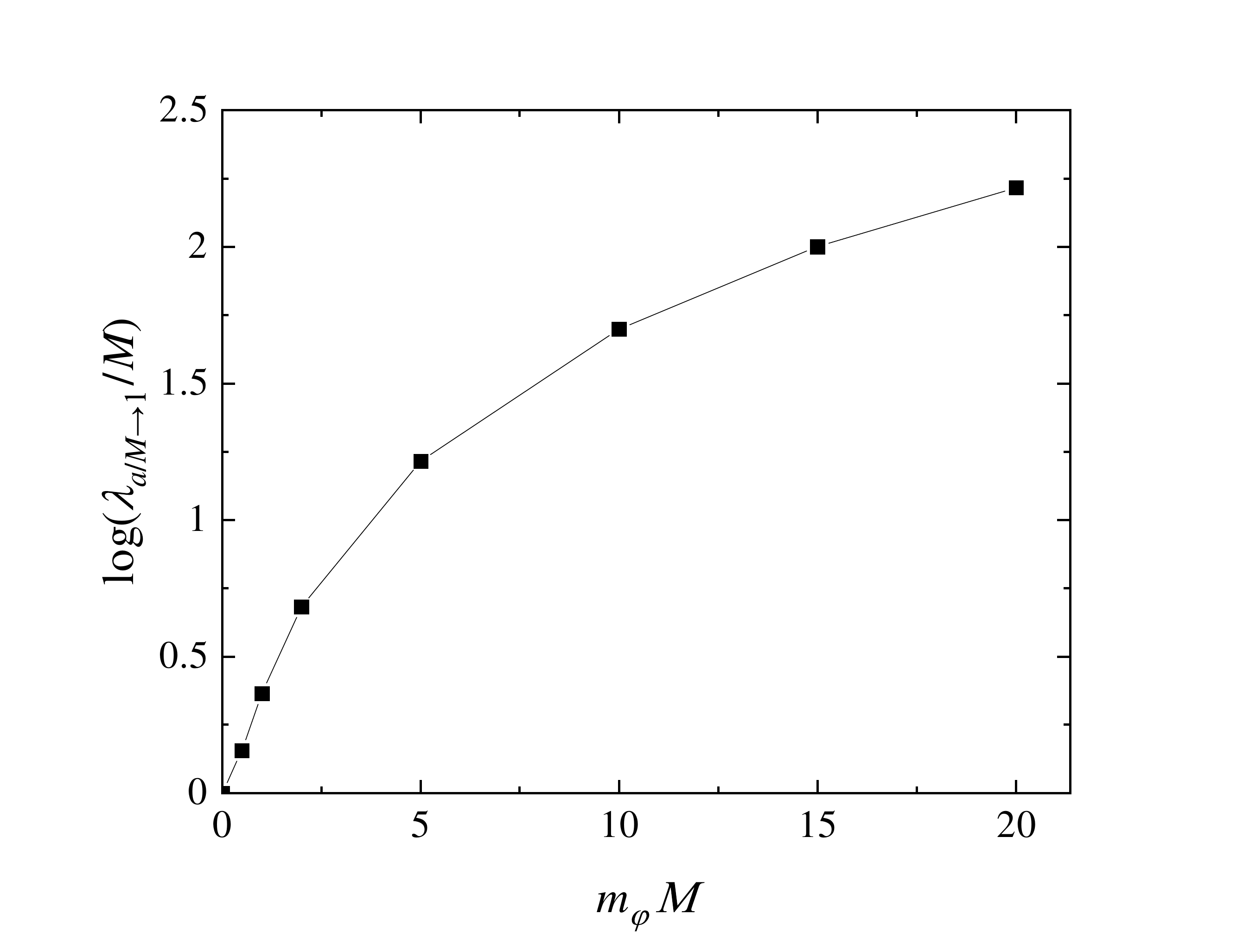}
		\caption{(\textit{left}) The instability lines, i.e., the contour lines $a(\lambda)$ dividing the parameter space into stable and unstable regions, for several scalar field masses $m_\varphi$.  (\textit{right}) The minimum value of the Gauss-Bonnet coupling parameter, denoted as $\lambda_{a/M \rightarrow 1}$, for which the instability can develop in the $a/M \rightarrow 1$ limit, as a function of the scalar field mass.}
		\label{fig:crit_line}
	\end{figure}

	Let us now examine more closely the effect that $m_\varphi$ has on the instability region of the Kerr black hole within EsGB gravity. Contour plots of the mode growth time, as a function of the angular momentum $a$ and the Gauss-Bonnet coupling parameter $\lambda$, are plotted in Fig. \ref{fig:contour} for the $m=0$ mode and several values of $m_\varphi$.\footnote{Modes with higher $m$ have normally larger growth times and develop instability for larger $\lambda$ \cite{Doneva:2020nbb} and that is why we will not focus on them here.} For better visibility, the critical lines where a change of stability is observed for different $m_\varphi$ is plotted in the left panel of Fig. \ref{fig:crit_line}.
	
	In accordance with the behavior of the signals presented in the previous subsection, the instability window where the Kerr solutions are unstable within the considered Gauss-Bonnet theory, shrinks with the increase of $m_\varphi$. More precisely, for fixed $a/M$, its left boundary is shifted to larger values of $\lambda$. This effect is very small if $m_\varphi M$ is of the order of one, which means that the Compton wavelength of the scalar field is comparable with the black hole horizon radius, but it can increase significantly for larger scalar field masses. If we pick a sufficiently large $\lambda$ (beyond the instability line), the growth times are weakly dependent on the particular value of $m_\varphi$. This is natural to expect since in this region the Gauss-Bonnet term prevails (due to the very large values of $\lambda$) and dictates the time evolution and the development of instability.
	
	The limiting value $\lambda_{a/M \rightarrow 1}$  for which the Kerr solution looses stability (and gives rise to new scalarized solution) in the $a/M \rightarrow 1$ limit, quickly increases with the increase of the mass as one can see in the right panel of Fig. \ref{fig:crit_line}. Again, the changes with respect to the massless case become significant only for large masses $m_\varphi M \gg 1$.

	\section{Conclusions}
	In the present paper, we have studied the development of spin-induced scalarization of the Kerr black hole within the framework of Gauss-Bonnet theory. In comparison to previous studies, we have focused on the impact of a nonzero scalar field potential. In particular, we studied the case where the potential corresponds to a scalar field mass term and neglected other terms, such as possible self-interaction. The reason for this is that we were mostly interested on the effect the potential will have on the onset of the instability and thus scalarization. While the self-interaction can have considerable effects on the nonlinear properties of the scalarized black holes, it does not influence the point of bifurcation.  Thus, for the purpose of the paper, the relevant Klein-Gordon equation for the scalar field, modified by an additional mass term, was evolved in (2+1) dimensions. The developed code has proven to be robust and well behaving in both stable and unstable region.

	As far as the time evolutions of stable models are concerned, one of the most significant changes due to the inclusion of nonzero scalar field mass are twofold: First, we observe the appearance of an oscillatory tail at late times, and second, if the scalar field mass is large enough so that the period of the oscillation in the tail is shorter than that of the quasinormal mode ringing itself, the oscillatory behavior due to the scalar field mass dominates the signal from very early times. The qualitative changes in the time evolution of unstable equilibrium models are smaller, though, compared to the  massless case, as soon as the exponential growth of the scalar field perturbation kicks in. 
	
	The scalar field mass has the effect of suppressing the scalar field itself and thus the instability -- the threshold value of the Gauss-Bonnet coupling constant above which instability develops, for fixed black hole angular momentum $a$, is shifted to larger values with increasing of $m_\varphi$.  Presented in a different way, if we fix the angular momentum $a$ and coupling constant $\lambda$, there exists a threshold $m_\varphi$ above which the scalar perturbations of the black hole stabilize. We have found that significant deviations in the region of instability are observed only for normalized scalar field masses of the order of one or greater,  which corresponds to the Compton wavelength of the scalar field being of the same order of the black hole horizon radius, or smaller. As far as the interior of the instability region is concerned, as soon as we move a bit further into the instability region, the growth times of the modes are only very weakly dependent on the scalar field mass. We point out that the threshold angular momentum for the development of the instability is not influenced by the presence of scalar field mass. This should be expected since the critical $a_{crit}/M$ is achieved in the $\lambda/M \rightarrow \infty$ limit. Clearly, in this case the mass term can not play any role. 
	
	Last, let us conclude with a few remarks on the astrophysical relevance of our study. It is well known that the lack of observational evidences of neutron star scalarization in close binary systems or in the inspiral phase before neutron star or black hole merger might lead to very strong constrains on the parameters of the theory predicting such a phenomenon. As a matter of fact, this is currently the case with the pure scalar-tensor theories and the Damour-Esposito-Farese model in particular \cite{Antoniadis2013,Freire2012,Shao:2017gwu}. A way to avoid this is to consider a mass of the scalar field which confines it to a radius not far away from the black hole and in this way suppresses the scalar dipole radiation. It is clear that spin-induced scalarization in EsGB gravity should exist also for neutron stars, including the case of differential rotation, which might significantly limit the allowed range of parameters after confronting with the observations. The inclusion of scalar field mass leads to suppression of the instability for smaller values of the Gauss-Bonnet coupling constant, but it still allows scalarized black holes to exist with angular momentum  as low as roughly $a/M=0.5$ (at least for large enough $\lambda/M$). Therefore, it might reconcile the theory with the observations  for a larger range of parameters, while still giving the opportunity to test the interesting phenomena of scalarization with future gravitational wave observations.

	\section*{Acknowledgements}
	DD and LC acknowledges financial support via an Emmy Noether Research Group funded by the German Research Foundation (DFG) under grant
	no. DO 1771/1-1. DD is indebted to the Baden-Wurttemberg Stiftung for the financial support of this research project by the Eliteprogramme for Postdocs.  SY would like to thank the University of Tuebingen for the financial support.  
	The partial support by the Bulgarian NSF Grant KP-06-H28/7 and the  Networking support by the COST Actions  CA16104 and CA16214 are also gratefully acknowledged. C.K. acknowledges support from the DFG reserach grant 	413873357.
	
	%%%%%%%%%%%%%%%%%%%%%%%%%%%%%%%%%%%%%%%%%%%%%%%%%%%%%%%%%%%%%%%%%%%%%%%%%%%%%%%
	
	\bibliographystyle{ieeetr}
	\bibliography{references,biblio_intro}

\end{document}